# Impurity-induced phonon thermal Hall effect in the antiferromagnetic phase of Sr$_2$IrO$_4$

A. Ataei[1], G. Grissonnanche[1], M-E. Boulanger[1], L. Chen[1], É. Lefrançois[1], V. Brouet[2], L. Taillefer[1,3]

*1 Institut quantique, Département de physique & RQMP, Université de Sherbrooke, Sherbrooke, Québec, Canada*

*2 Laboratoire de Physique des Solides, Université Paris-Saclay, CNRS, Orsay, France*

*3 Canadian Institute for Advanced Research, Toronto, Ontario, Canada*

**A thermal Hall effect is observed in an increasing number of insulators[1,2,3,4,5,6,7], often attributed to phonons[1,5,7,8,9,10,11], but the underlying mechanism is in most cases unknown. A coupling of phonons to spins has been invoked[12,13,14,15] and scattering of phonons by impurities or defects has been proposed[16,17,18,19,20], but there is no systematic evidence to support either scenario. Here we present a study on the effect of adding Rh impurities to the antiferromagnetic insulator Sr$_2$IrO$_4$, substituting for the spin-carrying Ir atoms[21,22]. We find that adding small concentrations of Rh impurities increases the thermal Hall conductivity $\kappa_{xy}$ dramatically, but adding enough Rh to suppress the magnetic order eventually decreases $\kappa_{xy}$ until it nearly vanishes. We conclude that the thermal Hall effect in this material is caused by the scattering of phonons by impurities embedded within a magnetic environment.**

The thermal Hall effect is used increasingly to probe insulators[23,24], materials with no mobile charge carriers. In the presence of a heat current $J$ along the $x$ axis and a magnetic field $H$ along the $z$ axis, a transverse temperature gradient $\nabla T$ (along the $y$ axis) can develop even if the carriers of heat are neutral (chargeless), provided they have chirality[25] or they acquire a handedness in the presence of a magnetic field. Of particular interest is the possibility that measurements of the thermal Hall conductivity $\kappa_{xy}$ could detect emergent excitations in quantum materials, such as Majorana fermions[26] or chiral magnons[27] – reportedly sighted in the spin liquid candidate α-RuCl$_3$ (refs. 3, 28).

Phonons are the dominant carriers of heat in all insulators, and so the first question to ask of any thermal Hall study is whether phonons are responsible for $\kappa_{xy}$. Initially, they were thought to generate only a very small thermal Hall effect, but we now know – from observations in multiferroic materials[7], cuprate Mott insulators[9,29], strontium titanate[5] and the antiferromagnetic insulator Cu$_3$TeO$_6$ (ref. 11), for example – that this is not true. However, although it is now clear that phonons can produce a sizable thermal Hall signal, the underlying microscopic mechanism is still not clear. Several theoretical scenarios have been proposed in the last few years[12,13,14,15,17,18,19,20,30], most recently focusing on the role played by impurity or defect scattering of phonons[17,18,19,20].
Here we report a systematic study of how Rh impurities affect the thermal Hall conductivity of the antiferromagnetic insulator Sr$_2$IrO$_4$. Despite the fact that Rh is isovalent to Ir, X-ray absorption experiments[31] have shown that, at small dopings, Rh adopts a valence 3+, with a non-magnetic 4d$^6$ configuration, different from the 5d$^5$ configuration of Ir$^{4+}$. Hence, it acts as a non-magnetic impurity, effectively trapping an electron and therefore doping the rest of the system with holes, as confirmed by ARPES experiments[32,33]. Magnetism is progressively suppressed and the system becomes increasingly metallic[34]. We find that 2% of Rh substituting for the spin-carrying Ir atoms (Fig. 1a) causes a 30-fold enhancement of the thermal Hall angle, $|\kappa_{xy}/\kappa_{xx}|$, while 15% of Rh – enough to suppress magnetic order (Fig. 1b) – brings this down to a negligible value. We conclude that impurities and magnetism both play a key role.

In Fig. 2, we show our data for the thermal conductivity $\kappa_{xx}$ and thermal Hall conductivity $\kappa_{xy}$ taken at a field of 15 T on five samples of $Sr_2Ir_{1-x}Rh_xO_4$, with Rh concentrations ranging from $x = 0$ to $x = 0.15$. We see that small concentrations of Rh, up to $x = 0.05$, yield only small variations in the magnitude of $\kappa_{xx}$ (Fig. 2a), no more than the factor 2-3 variation that is typical of the sample-to-sample variation seen in oxide crystals (see, for example, ref. 35). (Our $\kappa_{xx}$ data at $x = 0$ are similar to those previously reported for $Sr_2IrO_4$ (ref. 36); we attribute the difference in amplitude to a difference in crystalline (structural) quality.) By contrast, the same small concentrations of Rh cause a huge increase in the magnitude of $\kappa_{xy}$ (Fig. 2b). Plotting the ratio $|\kappa_{xy}/\kappa_{xx}|$ vs $T$ in Fig. 3a, we see that the peak value, at $T \approx 20$ K, increases very rapidly with $x$, at low $x$ (Fig. 3b). Specifically, $|\kappa_{xy}/\kappa_{xx}|$ is 30 times larger at $x = 0.02$ compared to $x = 0$, and 70 times at $x = 0.05$. This is compelling evidence that impurity scattering plays a key role in the mechanism responsible for the thermal Hall effect in this insulator.

The heat carriers at play in $Sr_2IrO_4$ are almost certainly phonons, given the striking similarity of $\kappa_{xy}(T)$ with the cuprate Mott insulators $La_2CuO_4$, $Nd_2CuO_4$ and $Sr_2CuCl_2O_2$ (ref. 29), materials in which phonons have been shown to cause $\kappa_{xy}$ (refs. 9, 35). Specifically, $|\kappa_{xy}(T)|$ peaks at the same temperature ($T \approx 20$ K) as the phonon-dominated $\kappa_{xx}(T)$, as in the cuprates – and indeed $SrTiO_3$ (ref. 5) and $Cu_3TeO_6$ (ref. 11), materials where phonons are also clearly the relevant heat carriers.

The heat carriers at $T = 20$ K, where $\kappa_{xy}$ is largest, are certainly not magnons because the gap in the magnon spectrum ensures that at such a temperature the contribution of magnons to heat transport is negligible[36]. Note also that charge carriers doped into the $IrO_2$ planes when Rh is added make a negligible contribution to $\kappa_{xx}$ and $\kappa_{xy}$, because of the very large electrical resistivity of our samples (see Methods).

Our first major finding is therefore this: impurity scattering plays a strong role in controlling the phonon thermal Hall effect in this material. We confirm this by introducing another type of impurity: La substituting for Sr. In Fig. 4, we report our data for four crystals of $Sr_{2-x}La_xIrO_4$, with La concentrations ranging from $x = 0$ to $x = 0.08$. (La doping in excess of $x = 0.10$ suppresses antiferromagnetic order.) We see that adding low levels of La impurities again causes an increase in $|\kappa_{xy}/\kappa_{xx}|$ (Fig. 4a). But the effect is much less pronounced than for Rh doping. Indeed, when measured at $T = 20$ K (and $H = 15$ T), $|\kappa_{xy}/\kappa_{xx}|$ reaches a maximal value that is 10 times smaller for La doping (Fig. 4b) compared to Rh doping (Fig. 3b). We infer that what matters is disorder on the spin-carrying site (Ir). In other words, spin also plays a key role.

We find further support for this inference by looking at higher doping levels. In Fig. 2b, we see that when enough Rh is added to fully suppress the long-range antiferromagnetic order, namely $x = 0.15$ (so that the Néel temperature goes to zero; Fig. 1b), the magnitude of $\kappa_{xy}$ becomes very small and $|\kappa_{xy}/\kappa_{xx}|$ is back down to its low value at $x = 0$ (Fig. 3b). A similar decrease of $|\kappa_{xy}/\kappa_{xx}|$ at high $x$ is found for La doping (Fig. 4b). This strongly suggests that magnetism is a key ingredient. This is our second major finding.

In summary, our doping studies of the antiferromagnetic insulator $Sr_2IrO_4$ show that impurities can generate a large phonon thermal Hall effect, especially when these impurities are embedded in a magnetic environment. This goes along the lines of a recent theoretical proposal based on resonant impurity scattering of phonons in a magnetic insulator[20]. A scenario of phonons scattered by impurities in a magnetic environment may be relevant for the thermal Hall effect of several other materials, such as the cuprates[4,9,29,35], whose magnetic order and crystal structure are very similar to those of $Sr_2IrO_4$, but also the cubic antiferromagnet $Cu_3TeO_6$ (ref. 11) and possibly spin liquid candidates such as the layered antiferromagnet α-RuCl$_3$ (refs. 3, 10).

**Acknowledgements.** L.T. acknowledges support from the Canadian Institute for Advanced Research (CIFAR) as a Fellow and funding from the Natural Sciences and Engineering Research Council of Canada (NSERC; PIN: 123817), the Fonds de recherche du Québec - Nature et Technologies (FRQNT), the Canada Foundation for Innovation (CFI), and a Canada Research Chair. This research was undertaken thanks in part to funding from the Canada First Research Excellence Fund. A.A. acknowledges support from the NSERC-CREATE program QSciTech and Bourse d'excellence de l'Institut quantique à l'Université de Sherbrooke.

**Author contributions.** A.A., G.G., M.E.B., L.C. and E.L. performed the thermal Hall conductivity measurements. A.A. prepared and characterized the samples. V.B. grew the single crystals of Sr$_2$IrO$_4$, Sr$_2$Ir$_{1-x}$Rh$_x$O$_4$ and Sr$_{2-x}$La$_x$IrO$_4$. A.A. and L.T. wrote the manuscript, in consultation with all authors. L.T. supervised the project.

**Author information.** The authors declare no competing financial interest. Correspondence and requests for materials should be addressed to A.A. (amirreza.ataei@usherbrooke.ca) or L.T. (louis.taillefer@usherbrooke.ca).

# METHODS

SAMPLES

**Sr$_2$IrO$_4$ and Rh-doped Sr$_2$IrO$_4$**. The single crystals of Sr$_2$IrO$_4$ and Sr$_2$Ir$_{1-x}$Rh$_x$O$_4$ were grown at the Université Paris-Saclay using a flux-grown technique[37], with Rh concentrations $x$ = 0.02, 0.05, 0.07 and 0.15. Contacts were made using silver paste and thin silver wires.

**La-doped Sr$_2$IrO$_4$**. The single crystals of Sr$_{2-x}$La$_x$IrO$_4$ were also grown at the Université Paris-Saclay with the same flux-grown technique, with La concentrations $x$ = 0.02, 0.04 and 0.08. Contacts were made in the same way.

THERMAL HALL MEASUREMENT

Thermal conductivity and thermal Hall conductivity measurements were performed with the standard one-heater-two-thermometer technique, as described in ref. 9. The thermal Hall conductivity is defined as $\kappa_{xy} = - \kappa_{yy} (\Delta T_y / \Delta T_x) (l / w)$, where $\kappa_{yy} = \kappa_{xx}$ in this material given its tetragonal crystal structure, and $\Delta T_y$ and $\Delta T_x$ are the transverse and longitudinal temperature differences across the sample, $l$ is the distance between the longitudinal contacts, and $w$ and $t$ are the sample width and thickness, respectively

In all measurements, the heat current $J$ was applied within the IrO$_2$ planes ($J // a$) and the magnetic field $H$ was applied normal to the IrO$_2$ planes ($H // c$).

ELECTRONIC THERMAL HALL CONDUCTIVITY

Adding Rh to Sr$_2$IrO$_4$ introduces electronic charge carriers which themselves contribute to thermal transport, in both $\kappa_{xx}$ and $\kappa_{xy}$. However, this contribution is negligible because all our Rh-doped samples have $\rho_{xx} > 0.6$ mΩ cm below 100 K (ref. 32). Indeed, in our sample with $x$ = 0.15, where the electrical Hall conductivity $\sigma_{xy}$ is largest (being the most metallic), its value is such that $L_0 \sigma_{xy} = 3$ μW / K$^2$ m for $H$ = 15 T at $T$ = 20 K (refs. 32, 38), only 1% of the measured value of $\kappa_{xy} / T$ for $H$ = 15 T at $T$ = 20 K. This negligible electronic contribution to $\kappa_{xy}$ becomes even smaller for $x < 0.15$.

**Data availability.** All the data that support the plots within this paper and other findings of this study are available from the corresponding authors upon reasonable request.

**MAIN FIGURES**

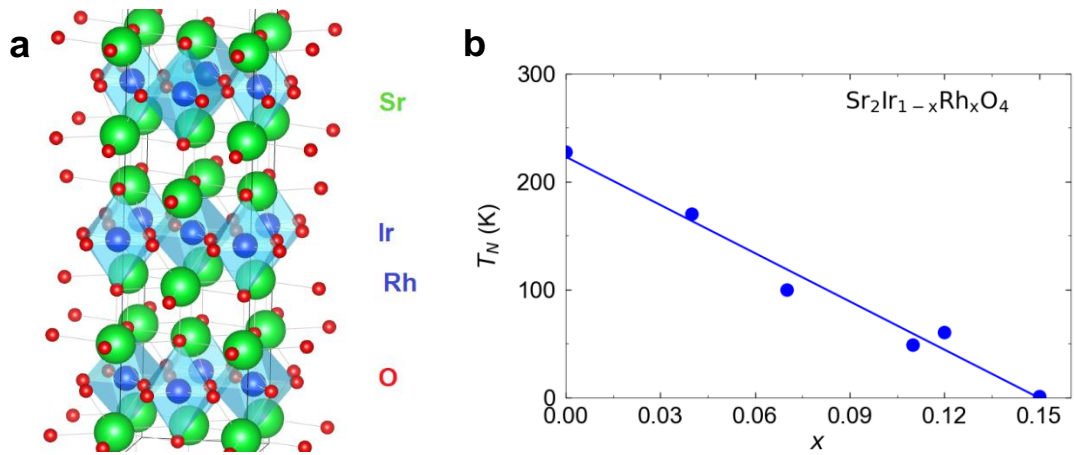

**Fig. 1 | Crystal structure and magnetic phase diagram of Rh-doped Sr$_2$IrO$_4$.**

**a)** Crystal structure of Sr$_2$IrO$_4$, showing the stacking of IrO$_2$ layers. The spins (moments) reside on the Ir sites, and order into a Néel antiferromagnetic state at low temperature. Rh impurities substitute for the Ir atoms; La impurities substitute for the Sr atoms. **b)** Temperature-doping phase diagram of Sr$_2$IrO$_4$, showing how the antiferromagnetic transition temperature $T_N$ (blue dots) decreases with Rh doping[32].

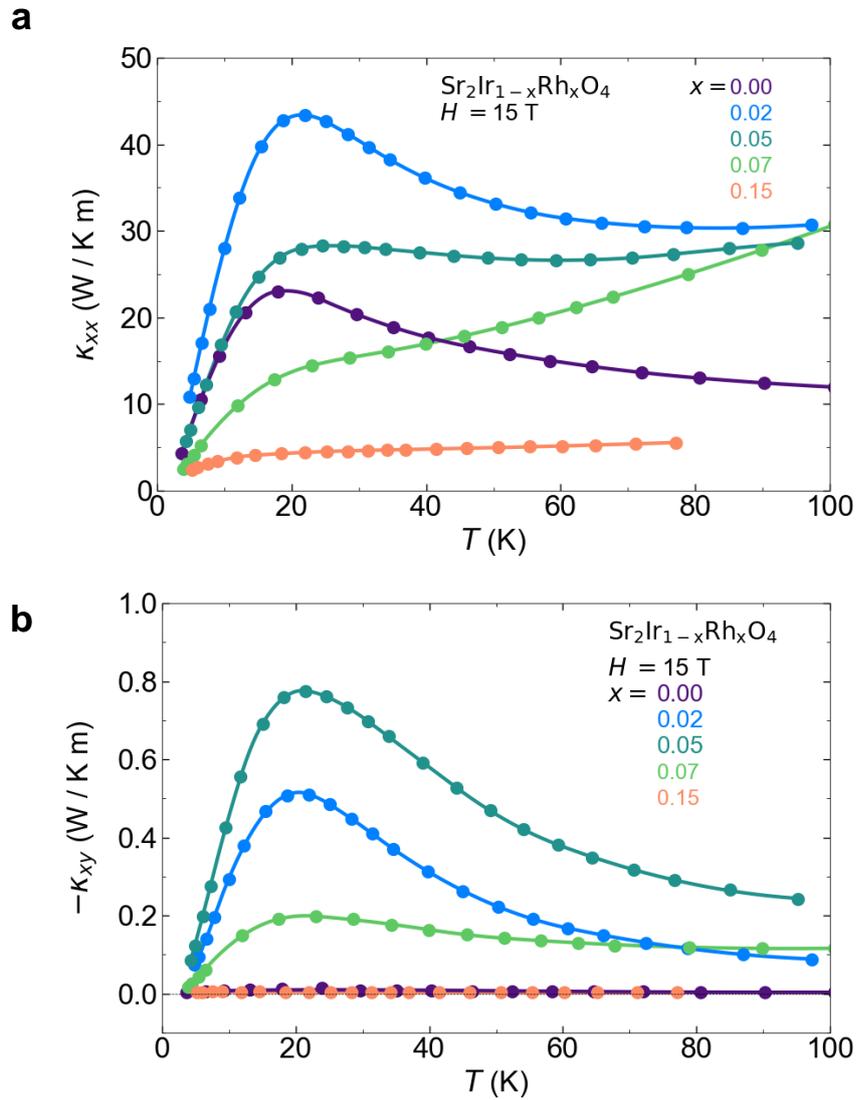

**Fig. 2 | Thermal conductivity and thermal Hall conductivity in Rh-doped Sr₂IrO₄.**

**a)** Thermal conductivity $\kappa_{xx}$ of $Sr_2Ir_{1-x}Rh_xO_4$ for a heat current parallel to the $IrO_2$ planes ($J // a // x$) and a magnetic field of 15 T applied normal to the planes ($H // c // z$), plotted as $\kappa_{xx}$ vs $T$, for $x = 0$ (violet), $x = 0.02$ (blue), $x = 0.05$ (dark green), $x = 0.07$ (light green), and $x = 0.15$ (pink). **b)** Thermal Hall conductivity $\kappa_{xy}$ for the same five samples, plotted as $-\kappa_{xy}$ vs $T$ ($\kappa_{xy}$ is negative in all samples at all temperatures).

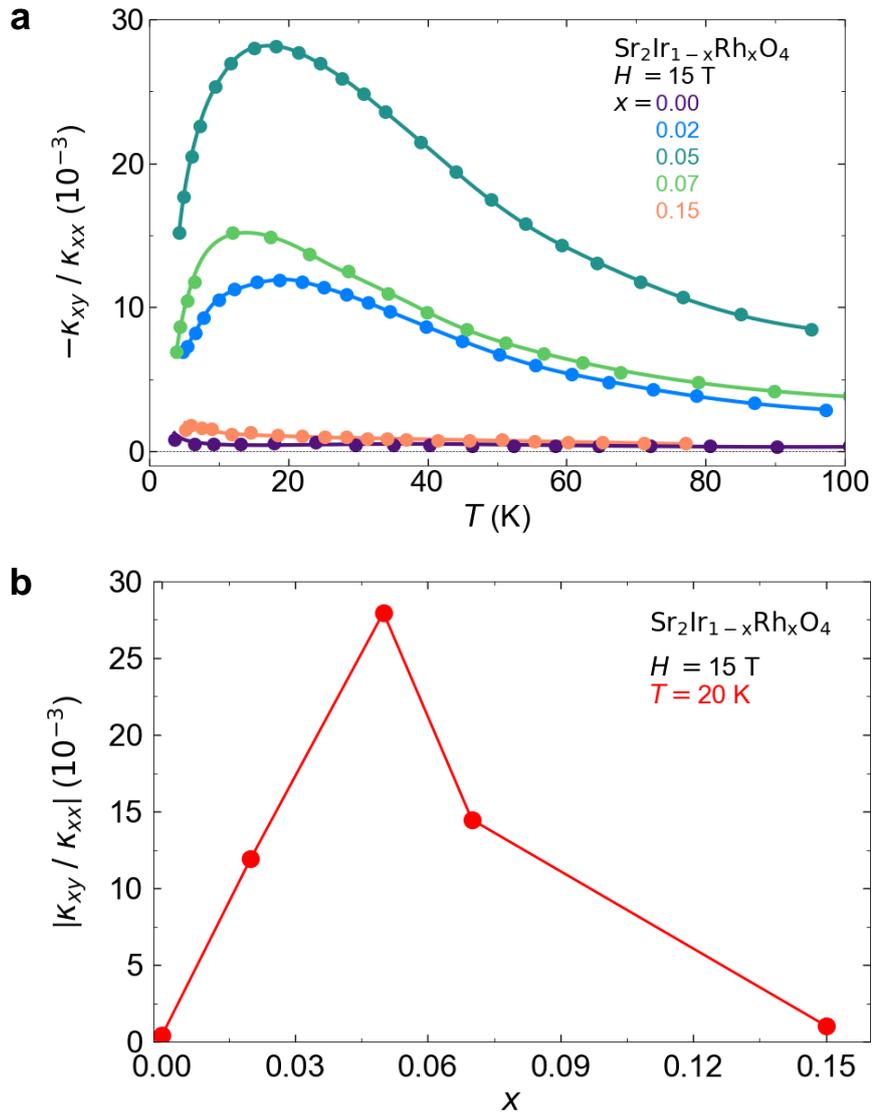

**Fig. 3 | Thermal Hall angle as a function of Rh doping.**

**a)** Thermal Hall angle in our five samples of $Sr_2Ir_{1-x}Rh_xO_4$, plotted as $|\kappa_{xy} / \kappa_{xx}|$ vs $T$, obtained from $\kappa_{xx}$ and $\kappa_{xy}$ data in Fig. 2. **b)** Magnitude of $|\kappa_{xy} / \kappa_{xx}|$, at $T = 20$ K, as a function of Rh doping. Note the 70-fold increase in $|\kappa_{xy} / \kappa_{xx}|$ between $x = 0$ and $x = 0.05$, and the subsequent decrease to a very small value at $x = 0.15$, where antiferromagnetic order has been suppressed (Fig. 1b).

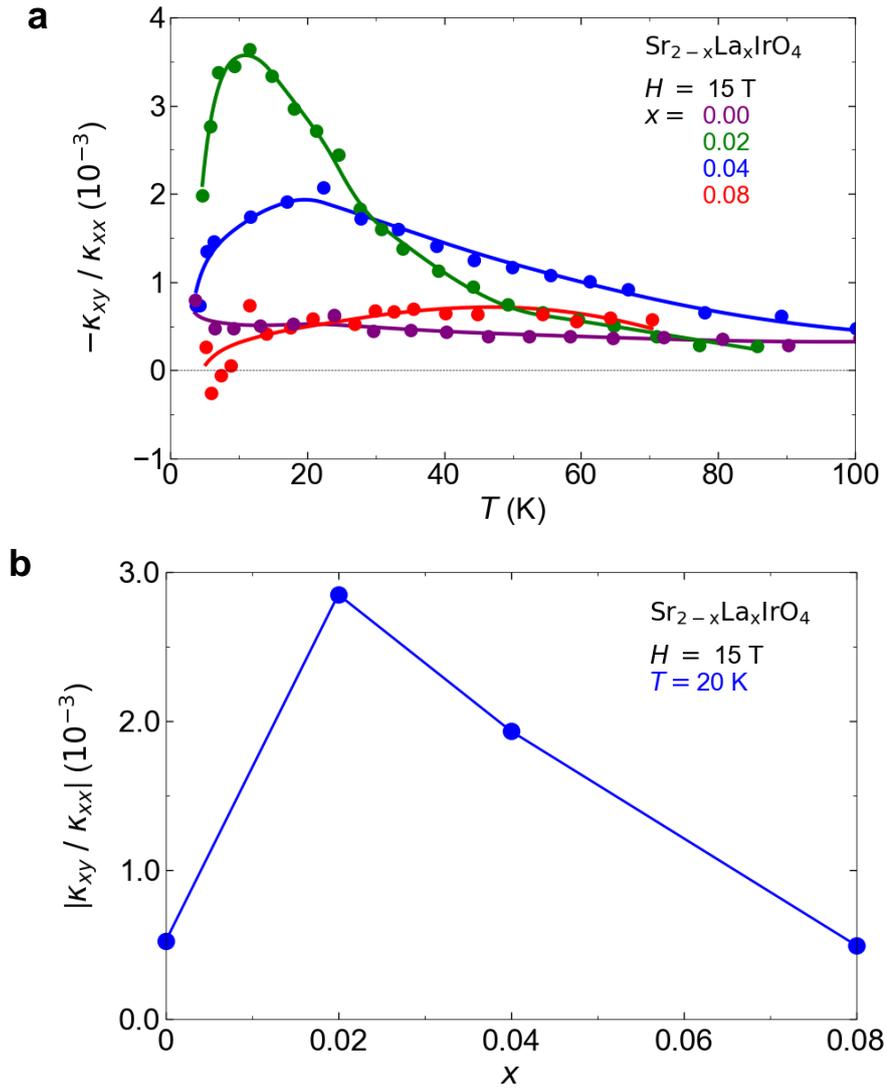

**Fig. 4 | Thermal Hall angle as a function of La doping.**

**a)** Thermal Hall angle in our four samples of $Sr_{2-x}La_xIrO_4$, plotted as $|\kappa_{xy}/\kappa_{xx}|$ vs $T$, obtained from $\kappa_{xx}$ and $\kappa_{xy}$ data in Extended Data Fig. 1. **b)** Magnitude of $|\kappa_{xy}/\kappa_{xx}|$, at $T = 20$ K, as a function of La doping. Note the 6-fold increase in $|\kappa_{xy}/\kappa_{xx}|$ between $x = 0$ and $x = 0.02$, and the subsequent decrease as magnetic order is being suppressed.

# EXTENDED DATA

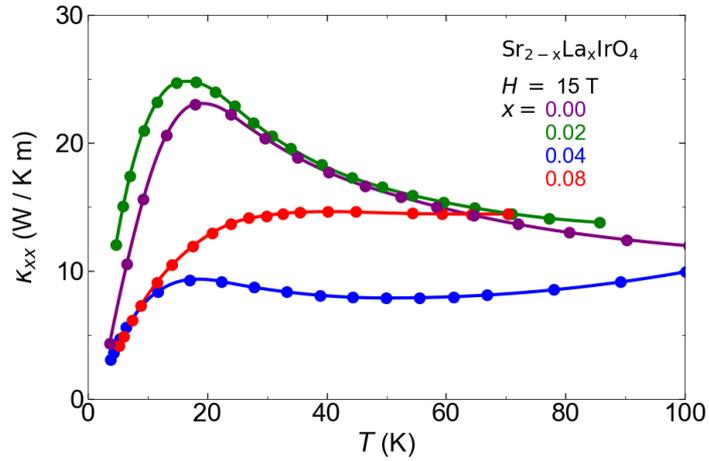

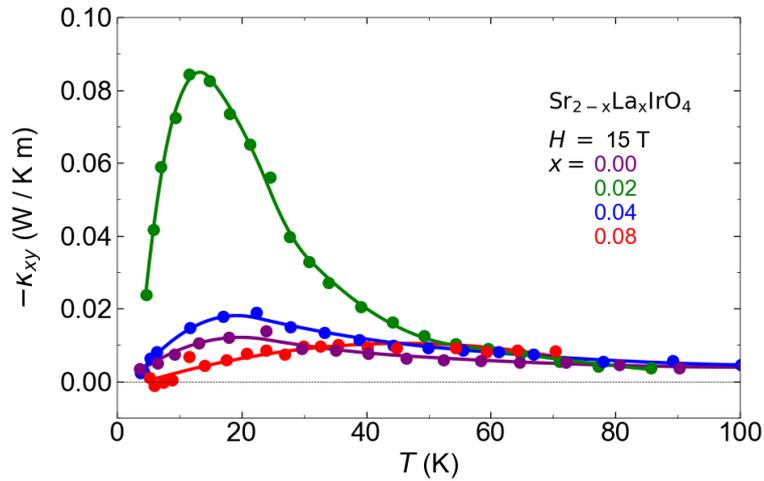

**Extended Data Fig. 1 | Thermal conductivities in La-doped $Sr_2IrO_4$.**

**a)** Thermal conductivity $\kappa_{xx}$ of $Sr_{2-x}La_xIrO_4$ for a heat current parallel to the $IrO_2$ planes ($J // a // x$) and a magnetic field of 15 T applied normal to the planes ($H // c // z$), plotted as $\kappa_{xx}$ vs $T$, for dopings $x$ as indicated. **b)** Thermal Hall conductivity $\kappa_{xy}$ for the same samples, plotted as $-\kappa_{xy}$ vs $T$ ($\kappa_{xy}$ is negative in all samples at all temperatures).